\documentclass[aps,twocolumn,showpacs]{revtex4}
\usepackage[dvips]{graphics}
\usepackage{color}
\definecolor{gold}{rgb}{0.85,0.66,0}
\definecolor{dblue}{rgb}{0,0,0.8}

\begin{document}

\title{{\textcolor{gold}{Current modulation at nano-scale level: A 
theoretical study}}}

\author{{\textcolor{dblue}{Santanu K. Maiti}}$^{1,2}$}

\affiliation{$^1$Theoretical Condensed Matter Physics Division, 
Saha Institute of Nuclear Physics, 1/AF, Bidhannagar, Kolkata-700 064, 
India \\
$^2$Department of Physics, Narasinha Dutt College, 129 Belilious Road, 
Howrah-711 101, India} 

\begin{abstract}
We explore the possibilities of current modulation at nano-scale level
using mesoscopic rings. A single mesoscopic ring or an array of such 
rings is used for current modulation where each ring is threaded by a 
time varying magnetic flux $\phi$ which plays the central role in the 
modulation action. Within a tight-binding framework, all the calculations 
are done based on the Green's function formalism. We present numerical 
results for the two-terminal conductance and current which support 
the essential features of current modulation. The analysis may be helpful 
in fabricating mesoscopic or nano-scale electronic devices.
\end{abstract}


\maketitle

\section{Introduction}

Electron transport in low-dimensional systems has drawn much attention 
in the field of theoretical as well as experimental research due to 
flourishing development in nanotechnology and nano-scale device modeling.
Low-dimensional model quantum systems are the basic building blocks for 
future generation of nano-electronic devices. Several exotic features 
are observed in this length scale owing to the effect of quantum 
interference. This effect is generally observed in samples with size 
much smaller or comparable to phase coherence length $L_{\phi}$, while 
the effect disappears for larger systems. A normal metal mesoscopic 
ring is a very good example to study the effect of quantum interference. 
Current trend of fabricating nano-scale devices has resulted much 
interest in characterization of ring type nanostructures. There are 
several methods for preparation of mesoscopic rings. For instance, gold 
rings can be designed using templates of suitable structure in combination 
with metal deposition through ion beam etching~\cite{hobb,pearson}. In 
a recent experiment, Yan {\it et al.} have proposed how gold rings can
be prepared by selective wetting of porous templates using polymer 
membranes~\cite{yan}. With such rings we can fabricate nano-scale 
electronic circuits which can be utilized for the operation of current 
modulation. To explore this phenomenon the ring is coupled to two 
electrodes, to form an electrode-ring-electrode bridge, where the ring 
is penetrated by a time varying magnetic flux $\phi$. Electron transport 
through a molecular bridge system was first studied theoretically by 
Aviram and Ratner~\cite{aviram} during $1970$'s. Following this pioneering 
work, several experiments have been done using different bridge systems 
to reveal the actual mechanism of electron transport. Though, to date a 
lot of theoretical~\cite{mag,lau,baer1,baer2,baer3,tagami,gold,cui,orella1,
orella2,fowler,peeters} as well as experimental works~\cite{reed1,reed2,
tali,fish} on two-terminal electron transport have been done addressing 
several important issues, yet the complete knowledge of conduction 
mechanism in nano-scale systems is still unclear to us. Transport 
properties are characterized by several significant factors like 
quantization of energy levels, quantum interference effect, 
ring-to-electrode interface geometry, etc. Furthermore, electron transport 
in the ring can also be modulated in other way by tuning the magnetic flux, 
the so-called Aharonov-Bohm (AB) flux, penetrated by the ring. 

Aim of the present paper is to illustrate the possibilities of 
current modulation at nano-scale level using simple mesoscopic rings. 
To achieve current modulation we design an electronic circuit using
a single mesoscopic ring or a cluster of such rings, where each ring
is penetrated by a time varying magnetic flux $\phi$ which plays the 
central role for the modulation action. For a constant DC voltage,
current through the circuit shows oscillatory behavior as a function 
of time $t$ depending on the phase of the magnetic flux $\phi$ passing
through the ring. Therefore, current modulation can be achieved simply
by tuning the phase of magnetic flux $\phi$ threaded by the ring. Within 
a tight-binding framework, a simple parametric approach~\cite{muj1,san3,
muj2,san1,sam,san2,hjo,walc1,walc2} is given and all the calculations are 
done through single particle Green's function technique to reveal the 
electron transport. Here we present numerical results for the two-terminal
conductance and current which clearly describe the essential features
of current modulation. Our exact analysis may be helpful for designing 
mesoscopic or nano-scale electronic devices. To the best of our knowledge 
the modulation action using such simple mesoscopic rings has not been 
addressed earlier in the literature.

The scheme of the present paper is as follows. With the brief introduction 
(Section I), in Section II, we describe the model and theoretical 
formulations for our calculations. Section III presents the significant 
results, and finally, we conclude our results in Section IV.

\section{Model and synopsis of the theoretical formulation} 

In the forthcoming two sub-sections we focus on two different circuit 
configurations, for our illustrative purposes, those are used for 
current modulation. Here we try to illustrate how a single mesoscopic 
ring or two such rings, where each ring is penetrated by a time varying
magnetic flux $\phi$, under a DC bias voltage can support an oscillating 
output current. A single mesoscopic ring can provide an oscillating
current with a particular frequency, while in the case of two rings,
oscillating currents with other frequencies can be obtained. These 
ideas may be generalized further to produce oscillating currents with
other frequencies by considering more number of rings.

\subsection{Circuit configuration I}

\begin{figure}[ht]
{\centering \resizebox*{7cm}{3.4cm}{\includegraphics{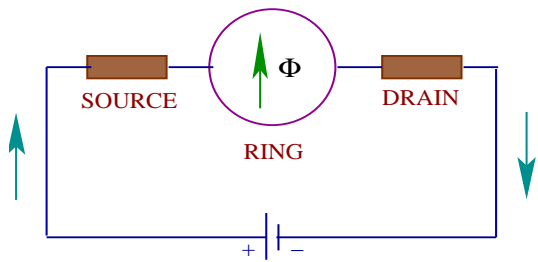}}\par}
\caption{(Color online). Actual scheme of connection with the battery
where a mesoscopic ring, subject to a time varying magnetic flux $\phi$,
is attached symmetrically to source and drain. The blue arrow indicates 
current direction in the circuit.}
\label{circuit1}
\end{figure}

Let us start by referring to Fig.~\ref{circuit1}. A mesoscopic ring, 
penetrated by a time varying magnetic flux $\phi$, is attached 
symmetrically to two semi-infinite one-dimensional ($1$D) metallic 
electrodes, namely, source and drain. These two electrodes are 
directly coupled to the positive and negative terminals of a battery,
a source of constant voltage. 
The time varying magnetic flux passing through the ring can be expressed
mathematically in the form,
\begin{equation}
\phi(t)=\frac{\phi_0}{2} \sin(\omega t)
\label{in1}
\end{equation}
where, $\phi_0=ch/e$ is the elementary flux-quantum, $\omega$ corresponds 
to the angular frequency and $t$ represents the time. This electronic
circuit provides an oscillating current in the output though a constant
DC input signal is applied which we will describe in the forthcoming
section. The frequency of the current is identical to that of the applied
flux $\phi(t)$.

\subsection{Circuit configuration II}

\begin{figure}[ht]
{\centering \resizebox*{7.5cm}{3.4cm}{\includegraphics{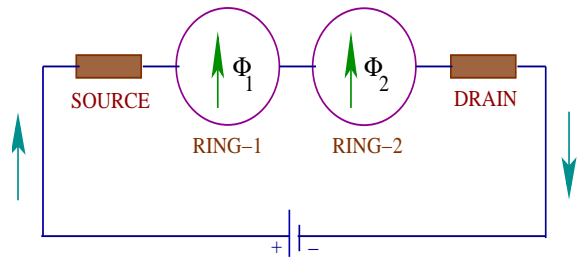}}\par}
\caption{(Color online). Actual scheme of connection with the battery 
where two mesoscopic rings, subject to time varying magnetic fluxes 
$\phi_1$ and $\phi_2$ are attached symmetrically to source and drain.
The blue arrow indicates current direction in the circuit.}
\label{circuit2}
\end{figure}

In Fig.~\ref{circuit2} two such mesoscopic rings those are directly 
coupled to each other, penetrated by time varying magnetic fluxes
$\phi_1$ and $\phi_2$ are attached symmetrically to the electrodes, viz, 
source and drain. A DC voltage source is connected to these two electrodes. 
The time varying magnetic fluxes are expressed mathematically as,
\begin{eqnarray}
\phi_1(t) & = & \frac{\phi_0}{2} \sin(\omega t) \label{in2} \\
\phi_2(t) & = & \frac{\phi_0}{2} \sin(\omega t + \delta) 
\label{in3}
\end{eqnarray}
where, $\delta$ refers to constant phase difference between the two fluxes
$\phi_1$ and $\phi_2$. Using this circuit configuration also oscillating
current in the output can be achieved, but in this case the frequency
of the current gets modified depending on the phase shift $\delta$.

\subsection{Theoretical formulation}

In this sub-section we will describe the basic theoretical formulation 
for calculation of conductance and current through a single mesoscopic
ring, penetrated by a magnetic flux $\phi$, attached to two source and 
drain. This similar theory is also used to study electron transport in 
an array of mesoscopic rings.

Using Landauer conductance formula~\cite{datta,marc} we determine
two-terminal conductance ($g$) of the mesoscopic ring. At much low 
temperatures and bias voltage it ($g$) can be written in the form,
\begin{equation}
g=\frac{2e^2}{h} T
\label{equ1}
\end{equation}
where, $T$ corresponds to the transmission probability of an electron 
across the ring. In terms of the Green's function of the ring and 
its coupling to two electrodes, the transmission probability can be 
expressed as~\cite{datta,marc},
\begin{equation}
T={\mbox{Tr}} \left[\Gamma_1 G_{R}^r \Gamma_2 G_{R}^a\right]
\label{equ2}
\end{equation}
where, $\Gamma_S$ and $\Gamma_D$ describe the coupling of the ring 
to the source and drain, respectively. Here, $G_R^r$ and $G_R^a$ 
are the retarded and advanced Green's functions, respectively, of the 
ring considering the effects of the electrodes. Now, for the full system 
i.e., the mesoscopic ring, source and drain, the Green's function is 
expressed as,
\begin{equation}
G=\left(E-H\right)^{-1}
\label{equ3}
\end{equation}
where, $E$ is the energy of the source electron. Evaluation of this 
Green's function needs the inversion of an infinite matrix, which is 
really a difficult task, since the full system consists of the finite 
size ring and two semi-infinite $1$D electrodes. However, the full 
system can be partitioned into sub-matrices corresponding to the 
individual sub-systems and the effective Green's function for the 
ring can be written in the form~\cite{marc,datta},
\begin{equation}
G_R=\left(E-H_R-\Sigma_S-\Sigma_D \right)^{-1}
\label{equ4}
\end{equation}
where, $H_R$ describes the Hamiltonian of the ring. Within the 
non-interacting picture, the tight-binding Hamiltonian of the ring 
can be expressed like,
\begin{equation}
H_R = \sum_i \epsilon_i c_i^{\dagger} c_i + \sum_{<ij>} v 
\left(c_i^{\dagger} c_j e^{i\theta}+ c_j^{\dagger} c_i e^{-i\theta}\right)
\label{equ5}
\end{equation}
where, $\epsilon_i$ and $v$ correspond to the site energy and 
nearest-neighbor hopping strength, respectively. $c_i^{\dagger}$ ($c_i$) 
is the creation (annihilation) operator of an electron at the site $i$
and $\theta=2 \pi \phi/N \phi_0$ is the phase factor due to the flux 
$\phi$ enclosed by the ring consists of $N$ atomic sites. A similar 
kind of tight-binding Hamiltonian is also used, except the phase factor 
$\theta$, to describe the electrodes where the Hamiltonian is parametrized 
by constant on-site potential $\epsilon^{\prime}$ and nearest-neighbor 
hopping integral $t^{\prime}$. The hopping integral between the ring and
source is $\tau_S$, while it is $\tau_D$ between the ring and drain. 
In Eq.~(\ref{equ4}), $\Sigma_S$ and $\Sigma_D$ are the self-energies 
due to the coupling of the ring to the source and drain, respectively, 
where all the information of the coupling are included into these 
self-energies.

To determine current, passing through the mesoscopic ring,
we use the expression~\cite{marc,datta},
\begin{equation}
I(V)=\frac{2 e}{h}\int \limits_{-\infty}^{\infty} 
\left(f_S-f_D\right) T(E)~ dE
\label{equ6}
\end{equation}
where, $f_{S(D)}=f\left(E-\mu_{S(D)}\right)$ gives the Fermi distribution
function with the electrochemical potential $\mu_{S(D)}=E_F\pm eV/2$ and
$E_F$ is the equilibrium Fermi energy. For the sake of simplicity,
we take the unit $c=e=h=1$ in our present calculations. 

\section{Numerical results and discussion}

To illustrate the numerical results, we begin our discussion by 
mentioning the values of different parameters used for our 
calculations. In the mesoscopic ring, the on-site energy $\epsilon_i$ 
is fixed to $0$ for all the atomic sites $i$ and nearest-neighbor 
hopping strength $v$ is set to $3$. While, for the side-attached 
electrodes the on-site energy ($\epsilon^{\prime}$) and 
nearest-neighbor hopping strength ($t^{\prime}$) are chosen as 
$0$ and $4$, respectively. The hopping strengths $\tau_S$ and 
$\tau_D$ are set as $\tau_S=\tau_D=2.5$. The equilibrium Fermi 
energy $E_F$ is fixed at $0$.

\subsection{Responses in circuit configuration I}

\begin{figure}[ht]
{\centering \resizebox*{7.8cm}{5cm}{\includegraphics{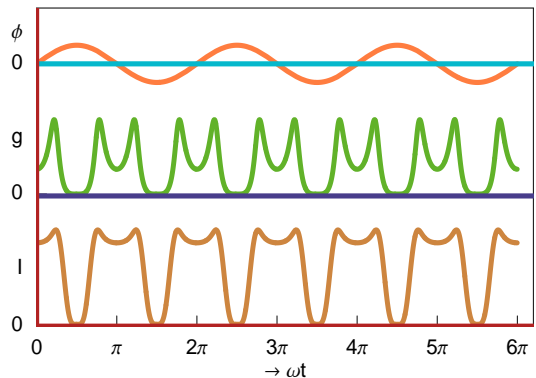}}\par}
\caption{(Color online). Responses in circuit configuration I. Upper,
middle and lower panels describe the time dependences of flux $\phi$,
conductance $g$ and current $I$ as a function of time $t$. Conductance 
is calculated at the energy $E=2.5$ and current is determined at the 
typical bias voltage $V=2.5$. The ring size is fixed at $N=20$. The
amplitudes are: $\phi_{max}=0.5$, $g_{max}=2$ and $I_{max}=2.54$.}
\label{current1}
\end{figure}

The modulation action for the circuit configuration I is clearly
illustrated in Fig.~\ref{current1}, where we compute all the results 
considering a ring with $N=20$. The upper panel presents the time 
variation of magnetic flux with amplitude $\phi_0/2$ whose mathematical 
form is given in Eq.~(\ref{in1}). The variation of conductance $g$ as 
a function of time $t$ is illustrated in the middle panel. Here we 
determine the typical conductance for a particular energy $E=2.5$. It 
shows that the conductance oscillates periodically as a function of 
$\omega t$ exhibiting $\pi$ periodicity and it gets the amplitude
$g_{max}=2$. This reveals that the transmission amplitude $T$ becomes 
unity since we get the relation $g=2T$ from the Landauer conductance
formula in our chosen unit $c=e=h=1$.
Now we try to justify the oscillating behavior of conductance with time
$t$. The probability amplitude of getting an electron from the source to
drain across the ring depends on the quantum interference effect of
the electronic waves passing through the upper and lower arms of the
ring. For a symmetrically connected ring (upper and lower arms are
identical to each other), penetrated by a magnetic flux $\phi$, the
probability amplitude of getting an electron across the ring becomes
exactly zero ($T=0$) for the typical flux, $\phi=\phi_0/2$. This 
vanishing behavior of transmission probability can be shown very 
easily by simple mathematical calculation as follows.

For a symmetrically connected ring, the wave functions passing through 
the upper and lower arms of the ring are given by,
\begin{eqnarray}
\psi_1 & = & \psi_0 e^{\frac{ie}{\hbar c} \int \limits_{\gamma_1} 
\vec{A}.\vec{dr}} \nonumber \\
\psi_2 & = & \psi_0 e^{\frac{ie}{\hbar c} \int \limits_{\gamma_2} 
\vec{A}.\vec{dr}} 
\label{equ10}
\end{eqnarray}
where, $\gamma_1$ and $\gamma_2$ are used to indicate the two different
paths of electron propagation along the two arms of the ring. $\psi_0$ 
denotes the wave function in absence of magnetic flux $\phi$ and it is 
same for both upper and lower arms as the ring is symmetrically coupled 
to the electrodes. $\vec{A}$ is the vector potential associated with the 
magnetic field $\vec{B}$ by the relation $\vec{B}= \vec{\nabla} \times 
\vec{A}$. Hence the probability amplitude of finding the electron passing 
through the ring can be calculated as,
\begin{equation}
|\psi_1 + \psi_2|^2 = 2|\psi_0|^2 + 2|\psi_0|^2 \cos \left({\frac{2\pi 
\phi}{\phi_0}}\right)
\label{equ11}
\end{equation}
where, $\phi = \oint \vec{A}.\vec{dr} = \int \int \vec{B}.\vec{ds}$
is the flux enclosed by the ring.

From Eq.~(\ref{equ11}) it is clearly observed that at $\phi=\phi_0/2$
the transmission probability of an electron drops exactly to zero. On
the other hand, for all other values of $\phi$ i.e., $\phi \ne \phi_0/2$,
electron transmission through the ring takes place which provides 
non-zero value of conductance. Thus, for the particular cases when 
$\phi(t)$ becomes maximum ($+ \phi_0/2$) or minimum ($- \phi_0/2$),
conductance drops to zero which is clearly shown from the conductance
spectrum (middle panel of Fig.~\ref{current1}). Hence, changing the
frequency of time dependent flux $\phi(t)$, periodicity in conductance
can be regulated. To visualize the oscillatory action more prominently 
we present the variation of current as a function of $\omega t$ in the 
lower panel of Fig.~\ref{current1}. The current $I$ through the ring 
is obtained by integrating over the transmission function $T$ (see 
Eq. (~\ref{equ6})). Here we compute the current for the typical bias
voltage $V=2.5$. Following the conductance pattern, the oscillatory 
behavior of the current is clearly understood, and like the conductance
spectrum current exhibits $\pi$ periodicity providing the amplitude
$I_{max}=2.54$. All these characteristic features suggest that an 
oscillatory response in the output is obtained though the ring is 
subject to a DC bias voltage.

\subsection{Responses in circuit configuration II}

\begin{figure}[ht]
{\centering \resizebox*{7.8cm}{5cm}{\includegraphics{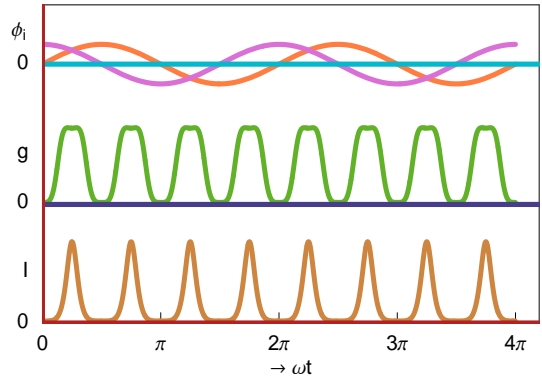}}\par}
\caption{(Color online). Responses in circuit configuration II. Upper,
middle and lower panels describe the time dependences of two fluxes 
$\phi_1$ (orange line) and $\phi_2$ (magenta line), conductance $g$ and 
current $I$ as a function of time $t$. Conductance is calculated at the 
energy $E=2.5$ and current is determined at the typical bias voltage 
$V=4$. In each ring, total number of atomic sites $N$ is fixed at $8$ and
we choose $\delta=\pi/2$. The amplitudes are: $\phi_{max}=0.5$, 
$g_{max}=1.9$ and $I_{max}=2$.}
\label{current2}
\end{figure}

Next, we concentrate on the responses obtained in the circuit 
configuration II. The results are illustrated in Fig.~\ref{current2},
where total number of atomic sites $N$ in each ring is fixed at $8$. 
In the upper panel, we plot the time dependent fluxes $\phi_1(t)$ 
(orange line) and $\phi_2(t)$ (magenta line) those pass through two 
different rings. A constant phase shift $\delta$ exists between these
two fluxes as mathematically expressed in Eqs.~(\ref{in2}) and 
(\ref{in3}). Here we set $\delta=\pi/2$. In the middle panel, we
describe the time dependence of conductance $g$ with amplitude 
$g_{max}=1.9$, where conductance is evaluated at the typical energy 
$E=2.5$. Conductance shows the oscillatory behavior as a function of 
$\omega t$ providing $\pi/2$ periodicity. Thus, for this circuit
configuration II, periodicity becomes exactly half compared to
the circuit configuration I. The explanation of $\pi/2$ periodicity
is as follows. For this two ring system, the transmission probability
depends on the combined effect of quantum interferences in the two 
rings. In ring-1, $\phi_1(t)$ is sinusoidal in form as described 
mathematically in Eq.~(\ref{in2}), while in ring-2, the variation
of flux $\phi_2(t)$ is the same as in ring-1 with a phase shift $\pi/2$.
Therefore, ring-1 and ring-2 enclose $\phi_0/2$ flux alternatively
in the interval $\omega t=\pi/2$, and accordingly, zero transmission
probability is achieved at this interval. In the same footing,
here we also describe the variation current $I$ with time $t$ (see
lower panel of Fig.~\ref{current2}) to support the oscillatory action 
observed in this circuit configuration II. The current is computed at 
the typical bias voltage $V=4$. The variation of current shows $\pi/2$
periodicity with an amplitude $I_{max}=2$ and this periodic nature
is well understood from the conductance spectrum. From these
conductance and current spectra it is manifested that in the two ring 
system which is subject to a DC bias voltage, the oscillatory response 
can be modulated very easily by tuning the phase difference $\delta$ 
between two time varying magnetic fluxes. 

Finally, we can say that extending this idea to an array of multi-ring
system in which different rings subject to time varying magnetic fluxes
in different phases, oscillatory responses can be achieved with $\pi/n$
frequencies, where $n$ corresponds to an integer. Our exact analysis may 
provide some significant insights in designing nano-electronic circuits.

\section{Concluding remarks}

In a nutshell, we have addressed the possibilities of current modulation
at nano-scale level using mesoscopic rings enclosing a time varying 
magnetic flux. We have shown that a single mesoscopic ring or two
such rings, subject to a DC bias voltage, can support an oscillating 
output current. A single mesoscopic ring can exhibit an oscillating 
current with a particular frequency associated with the flux $\phi(t)$,
while the frequency of the current can be regulated in the case of two
rings by tuning the phase difference $\delta$ between the fluxes
$\phi_1(t)$ and $\phi_2(t)$. The whole modulation action is based on 
the central idea of quantum interference effect in presence of flux 
$\phi$ in ring shaped geometries. We adopt a simple tight-binding 
framework to illustrate the model and all the calculations are done 
using single particle Green's function formalism. Our exact numerical 
results provide two-terminal conductance and current which clearly
describe the essential features of current modulation. Our analysis 
can be used in designing tailor made nano-scale electronic devices.

Throughout our work, we have described all the essential features
of current modulation for two different ring sizes. In circuit
configuration I, we have chosen a ring with total number of 
atomic sites $N=20$. On the other hand, in circuit configuration II, 
we have considered two identical rings, where each ring contains $8$ 
atomic sites. In our model calculations, these typical numbers ($20$
or $2\times8=16$) are chosen only for the sake of simplicity. Though the
results presented here change numerically with the ring size ($N$), but
all the basic features remain exactly invariant. To be more specific, it
is important to note that, in real situation the experimentally
achievable rings have typical diameters within the range $0.4$-$0.6$
$\mu$m. In such a small ring, unrealistically very high magnetic fields
are required to produce a quantum flux. To overcome this situation,
Hod {\em et al.} have studied extensively and proposed how to construct
nanometer scale devices, based on Aharonov-Bohm interferometry, those
can be operated in moderate magnetic fields~\cite{baer4,baer5,baer6,baer7}.

In the present paper we have done all the calculations by ignoring
the effects of the temperature, electron-electron correlation, etc. 
Due to these factors, any scattering process that appears in the
mesoscopic ring would have influence on electronic phases, and, in
consequences can disturb the quantum interference effects. Here we
have assumed that, in our sample all these effects are too small, and
accordingly, we have neglected all these factors in this particular 
study.

The importance of this article is mainly concerned with (i) the simplicity 
of the geometry and (ii) the smallness of the size.

\end{document}